\documentclass[%
 reprint,
 superscriptaddress,
 amsmath,amssymb,
 aps,
 prb,]{revtex4-2}

\usepackage{graphicx,dcolumn,bm,enumerate,eurosym}
\usepackage{amsfonts,amsmath,amsopn,amssymb}
\usepackage{graphicx,graphics}
\usepackage{color,verbatim,mathrsfs,braket,xargs,lipsum}
\usepackage[colorlinks=true, allcolors=blue]{hyperref}
\usepackage{mathtools}
\usepackage[dvipsnames]{xcolor}
\usepackage[colorinlistoftodos,prependcaption,textsize=tiny]{todonotes}
\newcommandx{\unsure}[2][1=]{\todo[linecolor=red,backgroundcolor=red!25,bordercolor=red,#1]{#2}}
\newcommandx{\change}[2][1=]{\todo[linecolor=blue,backgroundcolor=blue!25,bordercolor=blue,#1]{#2}}
\newcommandx{\info}[2][1=]{\todo[linecolor=OliveGreen,backgroundcolor=OliveGreen!25,bordercolor=OliveGreen,#1]{#2}}
\newcommandx{\improvement}[2][1=]{\todo[linecolor=Plum,backgroundcolor=Plum!25,bordercolor=Plum,#1]{#2}}
\newcommandx{\thiswillnotshow}[2][1=]{\todo[disable,#1]{#2}}
\usepackage[normalem]{ulem}

\hypersetup{
    colorlinks,
    linkcolor={blue!50!blue},
    citecolor={blue!50!blue},
    urlcolor={blue!80!black}
}

\usepackage[normalem]{ulem} 
\definecolor{mgreen}{RGB}{1,123,0}

\begin{document}

\preprint{APS/123-QED}

\title{Synchronization-Induced Limit Cycles and Hysteresis in Spinor Exciton-Polaritons}

\author{Saeid Asgarnezhad-Zorgabad}
\email{sasgarnezhad93@gmail.com}
\affiliation{Independent Researcher, Dublin, Ireland}

\author{Zahra Jalali-Mola}
\email{jalali.mola@gmail.com}
\affiliation{School of Physics and CRANN Institute, Trinity College Dublin, The University of Dublin, Dublin 2, D02 PN40, Ireland}

\author{Vladimir V. Konotop}
\email{vvkonotop@ciencias.ulisboa.pt}
\affiliation{Departamento de Física, Faculdade de Ciências, Universidade de Lisboa, Campo Grande 2, Edifício C8, Lisboa 1749-016, Portugal}

\author{Daniel Leykam}
\email{daniel\_leykam@sutd.edu.sg}
\affiliation{Science, Mathematics and Technology Cluster, Singapore University of Technology and Design,\\ 8 Somapah Road, 487372 Singapore}

\begin{abstract}
We study the nonlinear dynamics of spinor exciton-polaritons in resonantly pumped microcavities. Using steady-state analysis and computing eigenfrequencies via Bogoliubov-like linearization, we show that multiple Hopf bifurcations beyond conventional bistability occur within specific power regimes and are not associated with stable solutions of the system. Owing to nonlinear excitonic interactions, these transitions occur at different pump powers during forward and backward pump sweeps, leading to pronounced hysteresis between fixed-point and limit-cycle states. In the spinor regime, multiple limit-cycle branches and associated dynamical hysteresis loops emerge due to the interplay of polarization coupling and spin-dependent nonlinearities. We show how the locations of the bifurcation points can be tuned by the pump frequency and other system parameters. Our results identify synchronization as the organizing principle underlying limit-cycle formation, dynamical hysteresis, and bifurcation cascades in driven spinor exciton-polariton microcavities.
\end{abstract}

\date{\today}

\maketitle 

\section{Introduction}
The resonant interaction between Wannier-Mott excitons and cavity photons in optical microcavities produces quantum quasiparticles known as exciton-polaritons~\cite{RevModPhys.85.299}. Nonlinearities arising in this interaction under coherent optical pumping are important for a variety of applications ranging from superfluidity~\cite{RevModPhys.85.299,10.3788/PI.2022.R04} and condensation~\cite{doi:10.1126/science.1074464, ZJM2023, ZJM2026}, to optical bistability and limiting~\cite{Kavokin2017Microcavities,delteil2019towards}. Polaritonic optical bistability has been theoretically predicted~\cite{PhysRevA.54.3493} and experimentally verified~\cite{PhysRevLett.129.103601}, enabling simulation of classical Ising models for optimization~\cite{PhysRevB.99.195301}, logical gates~\cite{PhysRevB.82.033302}, optical switching~\cite{PhysRevLett.120.225301}, and many other applications~\cite{https://doi.org/10.1002/qute.202100137}. While early studies did not consider phase shifts due to the evolution of different transverse electric~(TE) and magnetic~(TM) fields in microcavity~(known as the TE-TM splitting~\cite{PhysRevLett.115.246401,PhysRevLett.112.066402}) and the spin degree of freedom in the exciton-exciton interaction, these effects are now known to play an important role in the formation of topological states and spin textures~\cite{https://doi.org/10.1002/qute.202100137}.

Optically pumped microcavities, in which cavity photons and excitons are linearly coupled while excitons interact nonlinearly, constitute a fundamental class of interacting exciton–polariton systems that can be quantitatively described in the lower and upper polariton basis~\cite{RevModPhys.85.299}. When upper polaritons are taken into account, higher cavity and excitonic states become accessible and, depending on the pump power and frequency detuning, the system may exhibit chaotic dynamics and complex pattern formation~\cite{PhysRevB.89.245307}. The polariton dynamics become further enriched when cavity polarization and exciton spin degrees of freedom are included~\cite{PhysRevLett.115.246401,PhysRevLett.112.066402}. In both cases of neglecting and considering spin and polarization degrees of freedom of cavity photons and exciton, the physical mechanism underlying the chaotic origin of the excited-state formation remains an open question. Despite the chaotic dynamics of this nonlinear interacting system, exciton and/or cavity-photon bistability—namely, the coexistence of two stable population states for fixed pump power and frequency—commensurate with Hopf bifurcation formation have already been predicted theoretically~\cite{PhysRevB.77.125324} and confirmed experimentally~\cite{10.1063/5.0136380,10.1063/5.0121316}.

In this paper, we show that resonantly driven spinor exciton-polaritons exhibit a hierarchy of synchronization-induced dynamical transitions beyond conventional optical bistability. Multiple Hopf bifurcations give rise to stable limit-cycle oscillations and hysteresis between fixed-point and limit-cycle attractors. Through forward and backward pump power sweeps, we uncover dynamical hysteresis loops associated with the creation and annihilation of these oscillatory states. Although the critical powers depend on pump frequency and other control parameters, synchronization between excitonic and photonic modes emerges as the universal mechanism governing limit-cycle formation and hysteretic transitions.

\section{Model}

\begin{figure}
    \centering
    \includegraphics[width=1\linewidth]{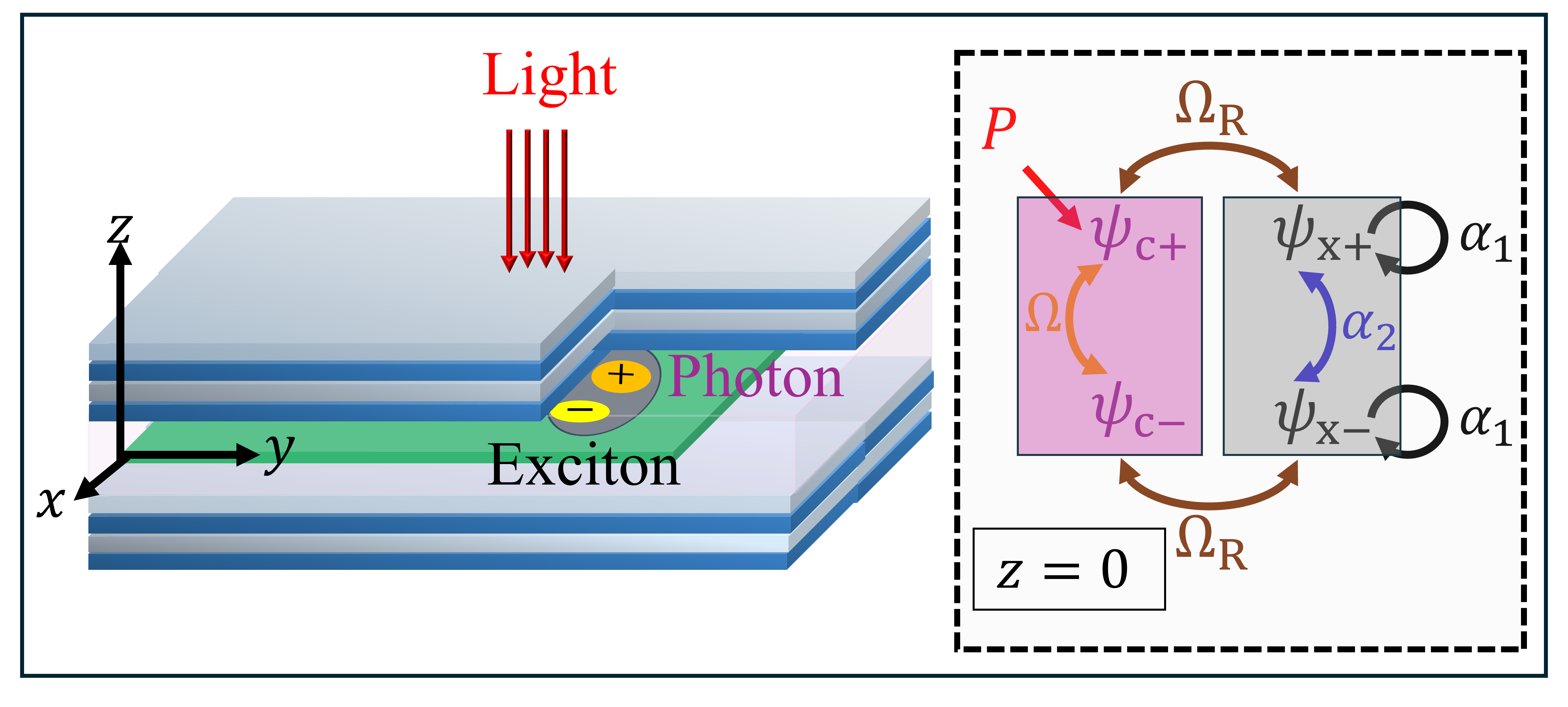}
    \caption{(a) Schematic of microcavity exciton–polariton system driven by circularly polarized light. Distributed Bragg mirrors with a central defect layer exhibiting linear polarization splitting $\Omega$, while the excitonic layer at $z=0$ hosts delocalized electron–hole pairs. Inset: cavity polarizations $\psi_{\mathrm{c}\pm}$, spinor excitons $\psi_{\mathrm{x}\pm}$, Rabi coupling $\Omega_{\mathrm R}$, and inter~(intra) exciton–exciton interactions $\alpha_1$~($\alpha_2$). }
    \label{Fig_Model_PL}
\end{figure}

A sketch of the microcavity exciton-polariton system is shown in Fig.~\ref{Fig_Model_PL}(a). It consists of a distributed Bragg mirror with a central defect layer designed to maximize electromagnetic-field localization at the defect center~($z=0$), for which an excitonic material in a quantum well structure is placed. We work in a rotated frame with respect to the external field's probe frequency $\omega$. Assuming $\gamma_\mathrm{c\pm}$ and $\gamma_\mathrm{x\pm}$ as the cavity and exciton decay rates, we denote the left~(right) circularly polarized photon wavefunctions as $\psi_{\text{c}\jmath}(\bm{r},t)\exp[(\mathrm{i}\omega-\gamma_\mathrm{c}/2)t]$, and spin-up~(down) exciton wavefunction as $\psi_{\text{x}\jmath}(\bm{r},t)\exp[(\mathrm{i}\omega-\gamma_\mathrm{x}/2)t]$ where $\jmath\in\{-,+\}$. Furthermore, we consider the coherent \textit{near-resonant} and spatially homogeneous pump in circular basis as $P_\pm \exp(i\omega t)$~\cite{RevModPhys.85.299}. Considering longitudinal~(transverse) cavity photon mass as $\text{m}_\text{l}$~($\text{m}_\text{t}$), respectively, the relative mass $\text{m}_\text{r}$ and effective mass $\text{m}$ can be expressed as~\cite{PhysRevLett.112.066402} 
\begin{align}    \mathrm{m}_\mathrm{r}&=\frac{\text{m}_\text{l}\text{m}_\text{t}}{\text{m}_\text{l}-\text{m}_\text{t}},\\
\mathrm{m}&=\frac{\text{m}_\text{l}\text{m}_\text{t}}{\text{m}_\text{l}+\text{m}_\text{t}}.
\end{align}
We then represent the TE-TM splitting term as $\beta=\hbar^2/(2\mathrm{m}_\mathrm{r})$~\cite{PhysRevLett.112.066402}. Furthermore, we assume exciton nonlinearities as intra-spin~(inter-spin) interactions characterized by $\alpha_1$~($\alpha_2$)~\cite{RevModPhys.85.299}. We include exciton and cavity photon decays through $\gamma_{\imath\pm}$ coefficient for $\imath\in\{\text{c},\text{x}\}$ describing cavity~(exciton) decay rates and assume $\gamma_{\imath+}=\gamma_{\imath-}=:\gamma_\imath$. Then we employ a single-particle Hamiltonian approximation for exciton-exciton interaction and obtain the equations governing the dynamics of the cavity and exciton wavefunctions as~\cite{PhysRevLett.112.066402} 
\begin{subequations}
\begin{align}
\label{Eq:Sudospin_Dynamics_cavity}
i\hbar \frac{\partial \psi_{\text{c}\pm}}{\partial t}
=& \Delta_{\rm c} \psi_{\text{c}\pm}
   - \frac{\hbar^2}{2\mathrm{m}}\nabla^2 \psi_{\text{c}\pm}  + \Omega_\text{R}\psi_{\text{x}\pm}
\nonumber \\
 & +\Bigl[\beta(\partial_y \mp \mathrm{i}\partial_x)^2
   - \frac{\Omega}{2} \Bigr] \psi_{\text{c}\mp}
   + P_\pm,  
\\
\label{Eq:Sudospin_Dynamics_exciton}
i\hbar \frac{\partial \psi_{\text{x}\pm}}{\partial t}
=& \Delta_\text{x} \psi_{\text{x}\pm}
   + \Omega_\text{R}\psi_{\text{c}\pm}+G(\psi_{\text{x}\pm})\psi_{\text{x}\pm},
\end{align}
\end{subequations}
where the effective cavity~(exciton) detuning is
\begin{equation}
   \Delta_{\imath}=\hbar\left(\omega_{\imath}-\omega-i\frac{\gamma_\imath}{2}\right), 
\end{equation}
$\omega_\imath$ is the bare cavity~(exciton) frequency, and
\begin{equation}
    G(\psi_{\text{x}\pm})=\alpha_1 \left|\psi_{\text{x}\pm}\right|^2
   + \alpha_2\left|\psi_{\text{x}\mp}\right|^2,
\end{equation} 
is the nonlinear term due to exciton-exciton interactions.

To explore exciton and cavity photon dynamics, we consider typical GaAs microcavity parameters; however, without loss of generality, we assume that the polarization and spin detunings and the decay rates have the same values and drop the $\pm$ signs. Throughout the text, we use meV and \(\mathrm{meV}/\mu\mathrm{m}\) to identify the physical units of the excitation frequency~[together with detunings] and pump power, respectively. Hence, cavity and exciton detunings are set as $\omega_\mathrm{c}=-2$ and, $\omega_\mathrm{x}=0$ while corresponding decay rates are $\gamma_\text{c}\approx66\times10^{-3}~$ and $\gamma_{\text{x}}=6.6\times10^{-3}~$), respectively, longitudinal~(transverse) photon mass as $\mathrm{m}_l=5\times10^{-5}\mathrm{m}_\text{e}$~($\mathrm{m}_t=10\mathrm{m}_l$), linear polarization splitting as $\Omega=0.2$~\cite{KLOPOTOWSKI2006511,PhysRevLett.112.066402}, exciton-polariton coupling as $\Omega_\text{R}=4$ and intra-spin interaction coefficient as $\alpha_1=6\times10^{-3}~$. 

\begin{figure}
    \centering
    \includegraphics[width=0.72\linewidth]{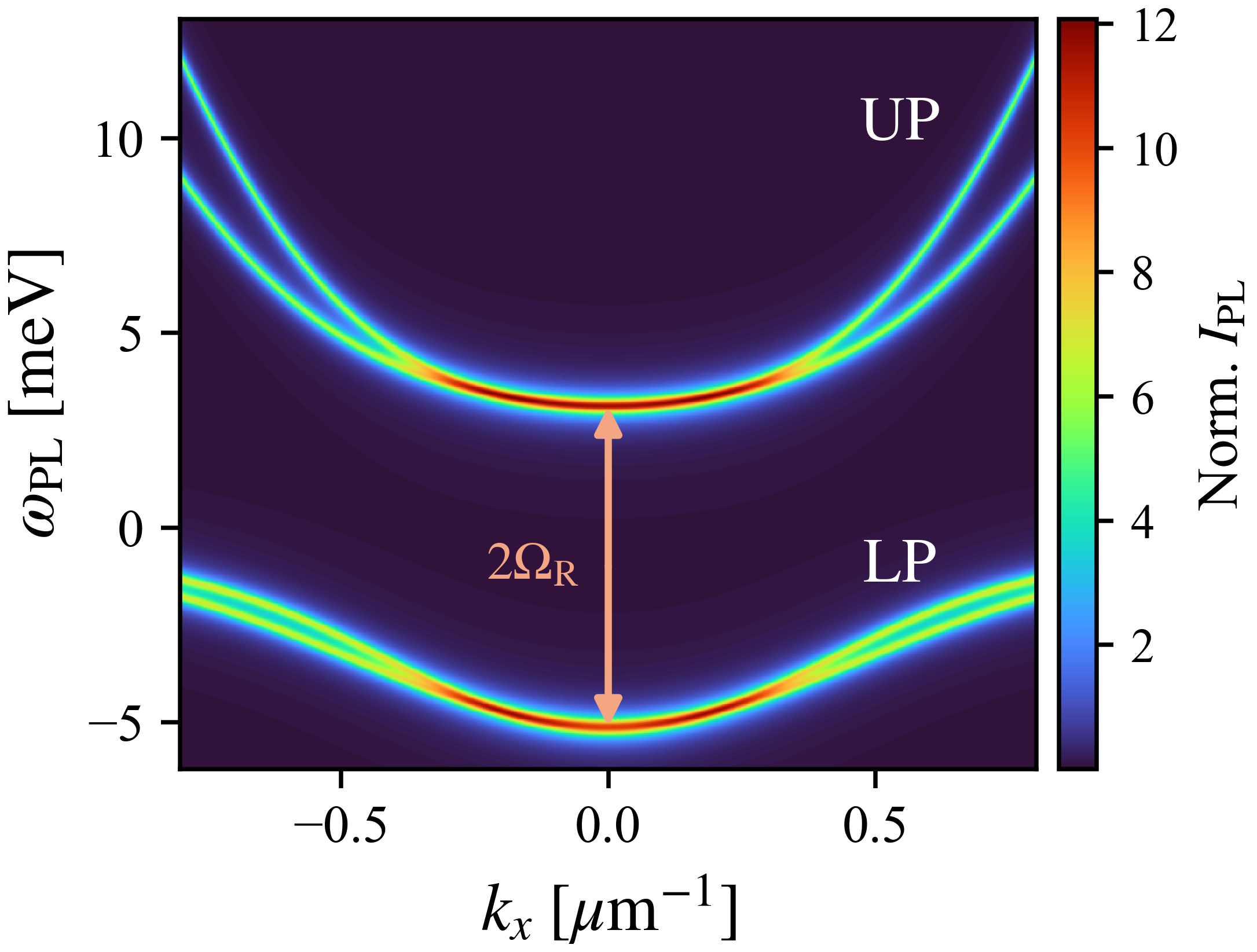}
    \caption{Photoluminescence dispersion showing lower~(LP) and upper~(UP) polariton branches with Rabi splitting $\Omega_{\mathrm R}=4$~meV for $\omega$ the eigenfrequency and $\bm{k}:=(k_x,k_z)$ the probe wavevector in momentum space. We assume a one-dimensional cavity and set $k_y=0$. We are interested in analyzing cavity and exciton population for frequencies close to the lower branch $\omega_1=-3$~meV and in the gap $\omega_2=2$~meV. See the text for other parameters.}
    \label{Fig_PL_Spectrum}
\end{figure}

To investigate the hybridization between exciton and cavity modes, we then calculate the photoluminescence dispersion by setting pump power and nonlinear terms to zero, namely, $\alpha_1=\alpha_2=0$ and $P_\pm=0$ in Eqs.~\eqref{Eq:Sudospin_Dynamics_cavity} and \eqref{Eq:Sudospin_Dynamics_exciton}, and calculating the eigenvalues of the linearized $4\times4$ Hamiltonian. In general, the photoluminescence spectrum implies four separate branches; however, for the parameters considered here, the weak polarization splitting produces nearly degenerate lower and upper polariton branches, shown in Fig.~\ref{Fig_PL_Spectrum}.  

The dispersion implies anti-crossing of the polaritonic branches due to the strong exciton-polariton hybridization~($\Omega_\text{R}\gg\gamma_\text{t}$). In the following, we investigate the time-steady-state and time-averaged dynamics, obtainable from Eqs.~\eqref{Eq:Sudospin_Dynamics_cavity} and \eqref{Eq:Sudospin_Dynamics_exciton}, for both two-component (neglecting the spin degree of freedom) and four-component (taking spin splitting into account) exciton-polariton evolutions. We focus on two frequencies: $\omega_1 = -3$, below the lower edge of the lower polariton branch, where the wavefunction is predominantly photonic, and $\omega_2 = 2$, in the gap between the lower and upper polariton branches, where both photonic and excitonic components are significant~\cite{RevModPhys.85.299}. 

\section{Stationary solutions}

We are interested in the stationary solutions of the homogeneous~(i.e., $\bm r$ independent) condensate under resonant pumping described by Eqs.~\eqref{Eq:Sudospin_Dynamics_cavity}  and \eqref{Eq:Sudospin_Dynamics_exciton}. Setting $\partial_x=\partial_y=\partial_t=0$ and introducing the four-component vector of the stationary solution $\bm \psi^{(\text{s})}=(\psi_{\text{c}+}^{\text{s}},\psi_{\text{c}-}^{\text{s}}, \psi_{\text{x}+}^{\text{s}},\psi_{\text{x}-}^{\text{s}})^T$ we require $\bm F(\bm \psi^{(\text{s})})=0$, where
\begin{align}
    \bm F=\left(\begin{matrix}
     \Delta_\text{c}\sigma_0 - \frac{\Omega}{2}\sigma_1e^{i\sigma_3\phi}& \Omega_{\rm R}\sigma_0\\
     \Omega_{\rm R}\sigma_0& \Delta_\text{x}\sigma_0 +G(\psi_{\text{x}\pm}^{\text{s}}) 
    \end{matrix}
    \right)\bm \psi^{\text{s}}+
     \left(
     \begin{matrix}
        P\bm{\mathrm{e}}_+ 
        \\
        \bm{0} 
     \end{matrix}
     \right),
     \label{Eq:Charac_Equation}
\end{align} 
and we have defined 
\begin{equation*}
    G(\psi_{\text{x}\pm})=\mathrm{diag}\left(G_+(\psi_{\text{x}\pm}),G_-(\psi_{\text{x}\pm})\right),
\end{equation*}
with $\bm{\mathrm{e}}_+:=(1,0)^\text{T}$ and $\bm{0}:=(0,0)^\text{T}$. The linear stability of a stationary state is then defined using the ansatz 
\begin{equation}
    \bm{\psi}(t)=\bm{\psi}^{\text{s}}+ \delta\bm{\psi}(t)
    \label{Eq:Linear_Perturb}
\end{equation}
with~($|\delta\bm{\psi}(t)|\ll|\bm{\psi}^{(\text{s})}|$)  which after the linearization of the system \eqref{Eq:Sudospin_Dynamics_cavity},  \eqref{Eq:Sudospin_Dynamics_exciton} yields the following standard eigenvalue problem~\cite{RevModPhys.63.991}
\begin{equation}
    \lambda  \bm{\phi}=\bm{J}\bm{\phi},
    \label{Eq:Jacobian}
\end{equation}
where the Jacobian matrix has entries ${J}_{ij}:=\partial F_{ij}/\partial\psi_j$. Stability and instability of the fixed point $\bm \psi^{(\text{s})}$ is identified by   $\text{Re}\, \lambda_P>0$ and $\text{Re}\,\lambda_P<0$, respectively. We then compute the eigenvalues of the $\bm J$ for both two-component and four-component exciton-polariton dynamics.

\subsection{Two-component limit}

For the outlined system parameters but with $\alpha_2=0$, and $\Omega=0$, Eqs.~\eqref{Eq:Sudospin_Dynamics_cavity} and \eqref{Eq:Sudospin_Dynamics_exciton} reduce to a two-component exciton-cavity system, where cavity polarization and exciton spin degrees of freedom are negligible. In this case, Eq.~\eqref{Eq:Charac_Equation} has an exact solution~\cite{PhysRevB.77.125324}. Considering only $\psi_{\text{x}+}$ component and dropping the subindex "$+$", the cavity wavefunction acquires the form $\psi_\text{c}^{\text{s}}=- (P+\Omega_\text{R}\psi_\text{x}^{\text{s}})/\Delta_\text{c}$ which upon substituting the exciton characteristic equation $(\Delta_\text{c}+\alpha_1|\psi_\text{x}^{\text{s}}|^2)\psi_\text{x}^{\text{s}}+\Omega_\text{R}\psi_\text{c}^{\text{s}}=0$ and defining $n_\text{x}=|\psi_\text{x}^{\text{s}}|^2$ and $\varpi_\mathrm{TC}:=\left(\Delta_\text{c}\Delta_\text{x}-\Omega_\text{R}^2\right)$, gives the following characteristic equation
\begin{equation}
\alpha_1^2\left|\Delta_\text{c}\right|^2n_\text{x}^3+2\alpha_1\mathrm{Re}\left[\Delta_\text{c}\varpi_\mathrm{TC}^*\right] n_\text{x}^2+\left|\varpi_\mathrm{TC}\right|^2n_\text{x}-(\Omega_\text{R}P)^2=0.
    \label{Eq:Exciton_Chara}
\end{equation}
Now the cavity population $n_\text{c}:=|\psi_\text{c}^{\text{s}}|$ is obtained as
\begin{equation}
    n_\text{c}=\frac{1}{\Omega_\text{R}^2}\left|\Delta_\text{x}+\alpha_1n_\text{x}\right|^2 n_\text{x}.
    \label{Eq:Cavity_Chara}
\end{equation}
Eq. (\ref{Eq:Exciton_Chara}) can be rewritten in a simpler form
\begin{align}    (\alpha_1\Delta_\text{c} n_\text{x} +\varpi_\mathrm{TC})^2 n_\text{x}-(\Omega_\text{R}P)^2=0.
\end{align}
We define $\eta=(\alpha_1\Delta_\text{c}/\varpi_\mathrm{TC}) n_\text{x}$ and $A=\alpha_1\Delta_\text{c}(\Omega_\text{R}P)^2/\varpi_\mathrm{TC}^3$ to obtain $(\eta+1)^2\eta-A=0$
whose discriminant is
\begin{align}
    \Delta=-A(27A+4).
    \label{Eq:Criteria}
\end{align} 
If $\Delta_c>0$, then $A>0$ and the equation has only one real root. Bistability emerges at $A=-4/27$. We note that $A$ is power-dependent, whose range depends on the frequency of the pump field~(see Eq.~\eqref{Eq:Critical_Power_Bistability}) for  $\omega = \omega_2$ and the power range satisfying \eqref{Eq:Criteria}.

\begin{figure}
    \centering
    \includegraphics[width=1\linewidth]{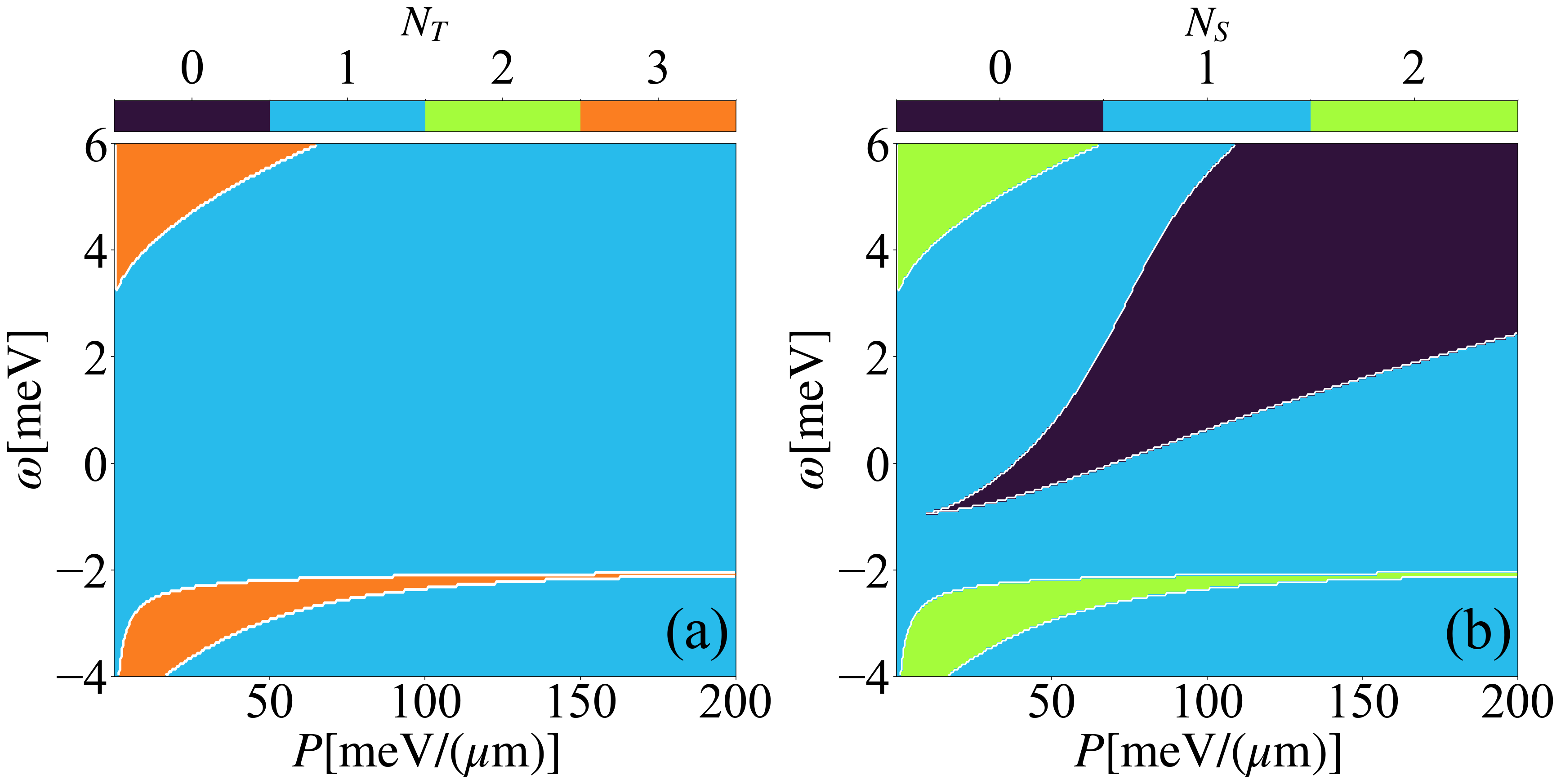}
    \caption{(a) Total number of solutions~($N_\text{T}$) and (b) number of stable solutions $N_\text{s}$~[right panel] as a function of pump frequency $\omega$ and pump power $P$ for two-component exciton polariton calculated numerically by the two-component version of the Eq.~\eqref{Eq:Jacobian}. Here we have used $\alpha_2=0$ and $\Omega=0$.}
    \label{fig:Stable_Two_component}
\end{figure}

Equations \eqref{Eq:Exciton_Chara} and \eqref{Eq:Cavity_Chara} are consistent with $\psi_\text{c(x)}^{\text{s}}$ satisfying $\bm{F}(\psi_\text{c(x)}^\text{s})=0$.
We then obtain $\bm{J}$ and corresponding eigenvalues by substituting \eqref{Eq:Linear_Perturb} into $\partial_t\bm{\psi}(t
)=\bm{F}(\bm{\psi}(t))$, performing a Taylor expansion up to first order of $\delta\bm{\psi}$, and making use of the stationary solution. We then find the total number of solutions $N_\text{T}$ and the number of stable solutions $N_\text{S}$, shown in Fig.~\ref{fig:Stable_Two_component}.

Near the lower polariton branch $\omega_1$, we observe bistability, for which \eqref{Eq:Exciton_Chara} represents lower and upper stable states and middle unstable states~\cite{PhysRevA.69.023809}, as clearly shown via the orange and green regions of Fig.~\ref{fig:Stable_Two_component} for $\omega<2$. In the mid-gap where both lower and upper polaritons contribute to the interaction, the system loses its stability, and all the solutions become unstable~(the dark blue region in Fig.~\ref{fig:Stable_Two_component}). This corresponds to a region of parametric instability in which nonlinear wave mixing leads to simultaneous excitation of the upper and lower polariton branches~\cite{PhysRevB.77.125324}. We finally recover the bistability for lower pump and high probe frequencies~(see green region of Fig.~\ref{fig:Stable_Two_component} with $\omega>3$ and $P<100$).

\begin{figure}
    \centering
    \includegraphics[width=1\linewidth]{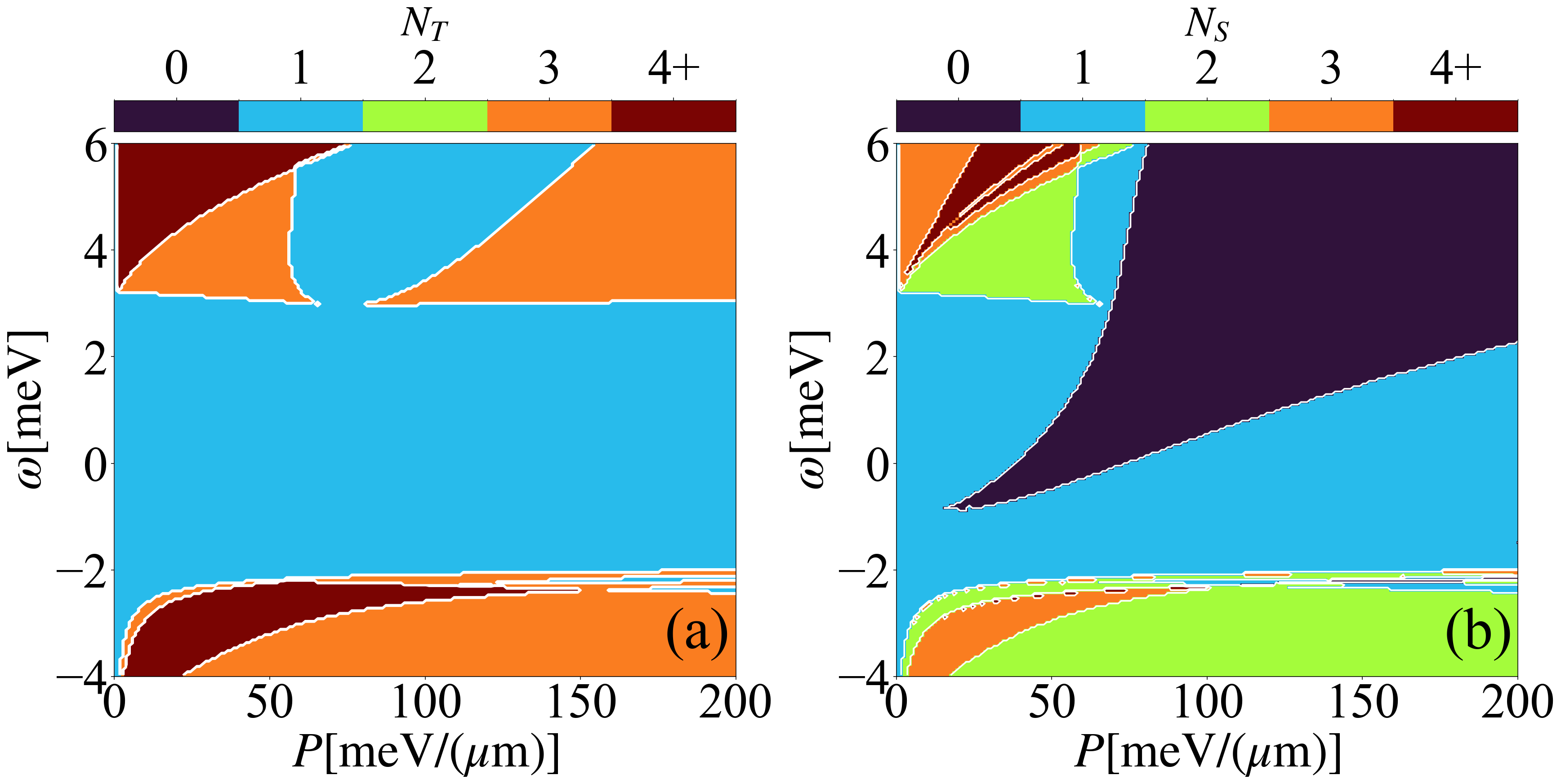}
    \caption{(a) Total number of solutions~($N_\text{T}$) and (b) number of stable solutions $N_\text{s}$ calculated numerically using Eq.~\eqref{Eq:Jacobian} as a function of pump frequency $\omega$ and pump power $P$ for four-component exciton polariton~(see analytical formulae Eqs.~\eqref{Eq:Charac_Exciton_Plus} and \eqref{Eq:Charac_Exciton_minus}). Here we have used $\alpha_2=-0.1\alpha_1$ and $\Omega=0.2$, see the text for the details.}
    \label{Fig_Steady_State}
\end{figure}

\subsection{Four-component model}

Inspired by the steady-state solution of the two-component limit, we now consider the four-component case described by Eq.~\eqref{Eq:Charac_Equation}. Defining $\bm \psi_\imath^\text{s}=(\psi_{\imath+}^\text{s},\psi_{\imath-}^\text{s})^\text{T}$, we obtain 
\begin{equation}
    \bm \psi_\text{c}^\text{s}=-\frac{\Delta_\text{c}\sigma_0+(\Omega/2)e^{\text{i}\sigma_3}\sigma_1}{\Delta_\text{c}^2+\left(\frac{\Omega}{2}\right)^2}\left(\Omega_\text{R}\bm \psi_\text{x}^\text{s}+P \mathrm{\bm{e}}_{+}\right).
    \label{Eq:Cavity_Four_Component}
\end{equation}
Upon substituting \eqref{Eq:Cavity_Four_Component} into $\Delta_\text{x}\sigma_0+\bm{G}(\psi_{\pm\text{x}}^\mathrm{s})$, and defining $n_\pm:=|\psi_{\pm\text{x}}^\mathrm{s}|^2$ $\mathcal{D}:=\Delta_\text{c}^2-(\Omega/2)^2$, $\varpi_\mathrm{FC}:=\Delta_\mathrm{x}\mathcal{D}-\Delta_\text{c}\Omega_\mathrm{R}^2$, $\alpha_+:=\alpha_1n_++\alpha_2n_-$ and $\alpha_-:=\alpha_1n_-+\alpha_2n_+$ we obtain amplitude of the exciton wavefunction in component form as
\begin{subequations}
\begin{align}
    \label{Eq:Charac_Exciton_Plus}
    \left(\varpi_\mathrm{FC}+\mathcal{D}\alpha_+\right)\psi_{\mathrm{x}+}-\frac{\Omega_\mathrm{R}^2\Omega}{2}e^{-i\phi}\psi_{\mathrm{x}-}&=\Delta_\text{c}\Omega_\mathrm{R}P,\\
    \label{Eq:Charac_Exciton_minus}
    \left(\varpi_\mathrm{FC}+\mathcal{D}\alpha_-\right)\psi_{\mathrm{x}-}-\frac{\Omega_\mathrm{R}^2\Omega}{2}e^{i\phi}\psi_{\mathrm{x}+}&=\frac{\Omega_\mathrm{R}^2\Omega}{2}e^{i\phi}P.
\end{align}
\end{subequations}

We observe that Eqs.~\eqref{Eq:Charac_Exciton_Plus} and \eqref{Eq:Charac_Exciton_minus} are nonlinear functions of the exciton populations $n_\pm$ and, dissimilar to \eqref{Eq:Exciton_Chara}, may contain higher powers of the exciton populations $n_\pm^{l>3}$. Thereby, by computing the eigenvalues of $\bm J$ and identifying the total number of stationary solutions $N_\text{T}$ and the number of stable stationary solutions $N_\text{S}$, we observe that for the frequencies close to the lower polariton branch $\omega_1$, or at higher probe frequencies but lower powers, the number of stable solutions can exceed two~(see the orange dark red colored regions in Fig.~\ref{Fig_Steady_State}). Thus, multistability can occur. 

The total number and number of stable solutions in Figs.~\ref{fig:Stable_Two_component} and~\ref{Fig_Steady_State} depend on the pump power $P$ and frequency $\omega$. Tuning these parameters and computing the eigenvalues of $\bm J$ we then identify saddle-node bifurcation~(the relevant eigenvalues remain negative) and Hopf bifurcations ($\mathrm{Re}[\lambda_{\text{max},P}]$ changes sign), denoted by SN and HB, respectively. Similar to the two component case, a region of parametric instability emerges in the gap between the lower and upper polariton branches.

\section{\label{Sec:Saddle_Hopf}Saddle node and Hopf bifurcations}

Next we address the long-time dynamics of the system, computing
time evolution of the cavity and exciton wavefunctions. In particular, considering $\gamma_\text{t}:=2\sum_{\imath}\gamma_{\imath}$ as the total decay rate of the system, we solve Eqs.~\eqref{Eq:Sudospin_Dynamics_cavity} and \eqref{Eq:Sudospin_Dynamics_exciton} within $t\in\left[0,T\right]$; for $T:=10^{3}\gamma_\text{t}^{-1}$, and perform time-averaging over the long-time dynamics of the cavity and exciton wavefunctions as $\braket{f}_T:=(2/T)\int_{T/2}^{T}dt f(t)$~\footnote{$T$ should be set long enough that the both $T/2$ and $T$ represents the long-time dynamics of the cavity and exciton population}. Considering $P_+:=P$ and $P_-=0$, we define the cavity~(exciton) population as 
\begin{equation}
    I_{\imath}(P):=\braket{|\psi_{\imath+}|^2+|\psi_{\imath-}|^2}_t. \label{Eq:Power_Population}
\end{equation} 
We use this equation to identify whether the system relaxes to a stable fixed point, enters a limit cycle, or follows a distinct history-dependent branch for backward pump variation. Numerically, this can be done by employing the pump continuation method, using the final state obtained at the previous pump power step as the initial condition for the next step. This continuation procedure allows tracking of history-dependent attractors and hysteretic branches. In what follows, we compute \eqref{Eq:Power_Population} for both two-component and four-component exciton-polariton systems.

\begin{figure}
    \centering
    \includegraphics[width=1\linewidth]{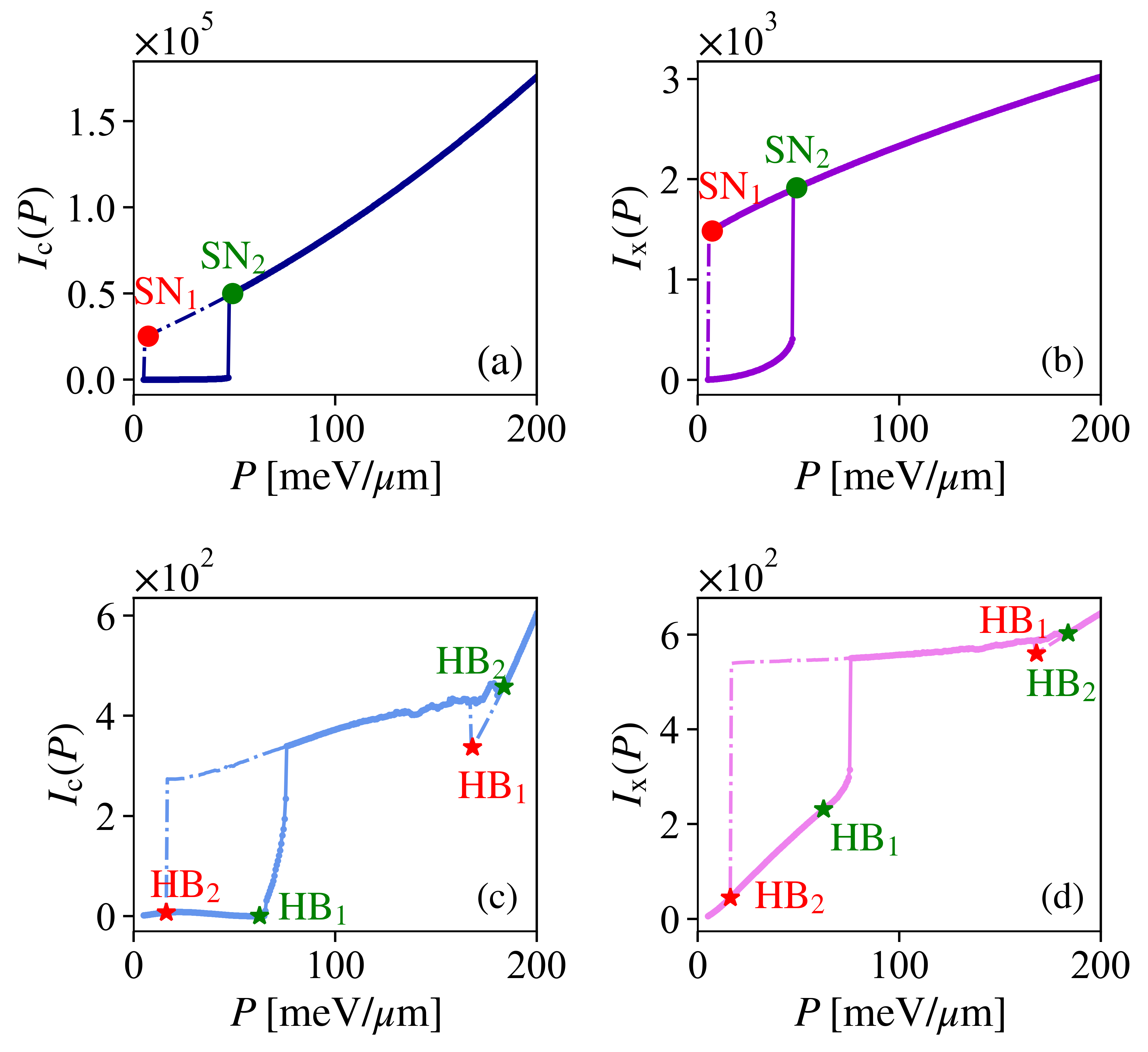}
    \caption{(a) Time-averaged cavity population $I_\mathrm{c}$ for both increasing~(dark-blue solid line) and decreasing~(dashed-dotted dark-blue line) pump power sweeps. (b) Time-averaged exciton population $I_\mathrm{x}$ for increasing~(dark-blue solid line) and decreasing~(dashed-dotted dark-blue line) pump power sweeps. For these panels, we have used $\omega_1=-3$, while the other parameters are the same as in Fig.~\ref{Fig_PL_Spectrum}. We characterize saddle-node(points) via $\text{SN}_{1(2)}$ of the increasing~(decreasing) pump with green~(red) dots, respectively. (c,d) Same as (a,b), but with $\omega_2 = 2$. Critical points $\mathrm{HB}_1$ and $\mathrm{HB}_2$ characterized by green~(red) stars denote Hopf bifurcations.}
    \label{Two_Component_Power_Population}
\end{figure}

\subsection{Two component limit}

We start our analysis by assuming a two-component problem~(namely, with $\alpha_2=0$ and $\Omega=0$). In this simpler case, critical points in the power spectrum~(exciton and cavity population for various pump powers) for specific probe frequency $\omega$ satisfy 
\begin{equation}
    3\alpha_1^2\left|\Delta_\mathrm{c}\right|^2n_\mathrm{x}^2+4\alpha_1\text{Re}[\Delta_\mathrm{c}\varpi_\text{TC}^*]n_\mathrm{x}+\left|\varpi_\text{TC}\right|^2=0.
    \label{Eq:Bistable_Charac}
\end{equation}
We now investigate the critical points for probe frequencies $\omega_1$ and $\omega_2$. For $\omega = \omega_1 =-3$ and the outlined system parameters, we obtain $\Delta_\mathrm{c}=1-0.033\mathrm{i}$, $\Delta_\mathrm{x}=3-0.003\mathrm{i}$ and $\varpi_\text{TC}=-13.0-0.103\mathrm{i}$~[in the units of (meV)$^2$]. Plugging these numerical values into \eqref{Eq:Bistable_Charac}, we obtain the critical points
\begin{equation}
    n_\mathrm{x,cr}^{(1)}=476.75, \quad n_\mathrm{x,cr}^{(2)}=1425.44,
\end{equation}
which upon substitution into Eq.~\eqref{Eq:Exciton_Chara} and using $P>0$, it follows that
\begin{equation}
    P_\mathrm{cr,1}=5.03,\quad P_\mathrm{cr,2}=47.30,
    \label{Eq:Critical_Power_Bistability}
\end{equation}
for $P_{\mathrm{cr},1}$~($P_{\mathrm{cr},2}$) being the decreasing~(increasing) power saddle node bifurcation, respectively. Furthermore, by calculating Jacobian matrix, we obtain $\max\text{Re}[\lambda_{P_\mathrm{cr,2}}]=-2.914\times10^{-14}$ and $\max\text{Re}[\lambda_{P_\mathrm{cr,1}}]=1.506\times10^{-12}$, indicating that these critical points correspond to a saddle-node bifurcation whose Jacobian eigenvalues satisfy $\max\text{Re}[\lambda_{\mathrm{SN}}]\to0$. For the mid-gap frequency $\omega = \omega_2=2$, we obtain $\Delta_\mathrm{c}=-3-0.033\mathrm{i}$, $\Delta_\mathrm{x}=-1-0.003\mathrm{i}$ and $\varpi_\text{TC}=-8.0-0.079\mathrm{i}$. Plugging these values into \eqref{Eq:Bistable_Charac}, we obtain two negative solutions for $n_\mathrm{x}$, indicating the destabilization of the saddle node. 

We then exploit Eq.~\eqref{Eq:Power_Population} to calculate time-averaged exciton and polariton populations using increasing- and decreasing-pump-power sweeps. To this aim, we solve \eqref{Eq:Sudospin_Dynamics_cavity} and \eqref{Eq:Sudospin_Dynamics_exciton} by assuming noise to exciton and cavity wavefunctions as the initial condition. For $\omega=\omega_1$ late-time dynamics follow different stable branches, see dark-blue and dark-violet lines in Fig.~\ref{Two_Component_Power_Population}(a) and \ref{Two_Component_Power_Population}(b), respectively. Specifically, we observe optical bistability for $P\approx47$~(green dot $\text{SN}_1$ in Fig.~\ref{Two_Component_Power_Population}(a) and \ref{Two_Component_Power_Population}(b)). For decreasing pump power, the cavity and exciton wavefunctions possess different memory due to the exciton-exciton interaction, and we obtain optical bistability at $P\approx7$~(red dot $\text{SN}_1$ in Fig.~\ref{Two_Component_Power_Population}(a) and \ref{Two_Component_Power_Population}(b)). These findings are in good agreement with the steady-state critical-power calculation~[see Eq.~\eqref{Eq:Critical_Power_Bistability}].

Applying Eq.~\eqref{Eq:Bistable_Charac} for $\omega=\omega_2$ yields negative values for $n_\mathrm{x}$, indicating that SN points lose stability. To investigate the system's evolution, we use Eq.~\eqref{Eq:Power_Population} to calculate exciton and cavity populations for both increasing and decreasing pump power sweeps. For increasing pump power, we see a non-chaotic increment of $I_\imath(P)$ up to $P\approx63$, after which cavity and exciton populations experience a jump to a higher state, evolving with near-plateau population amplitude~(up to $P\approx182$)~[see light-blue and magenta solid lines in Fig.~\ref{Two_Component_Power_Population}(c) and \ref{Two_Component_Power_Population}(d)]. Both populations evolve similarly to Fig.~\ref{Two_Component_Power_Population}(a) and \ref{Two_Component_Power_Population}(b) for $P>182$. For decreasing pump power, populations start to evolve as a semi-plateau for $P\approx167$, and this higher-state evolution lasts up to $P\approx16$, for which both cavity and exciton populations undergo a jump and relax to the ground state.   

To characterize the criticality behind this evolution, we calculate the eigenvalues of the $\bm J$ matrix: for increasing pump power and $P_\mathrm{F1}=62.475$~(green $\text{HB}_1$ star in Fig.~\ref{Two_Component_Power_Population}(c) and \ref{Two_Component_Power_Population}(d)) the eigenvalue changes from $\lambda_{P_\mathrm{F1}}=-2.6251\times10^{-5}+4.2252\mathrm{i}$ to $\lambda_{P_\mathrm{F1}}=4.5873\times10^{-5}+4.2290\mathrm{i}$, while for $P_\mathrm{F2}=182.134$~(green $\text{HB}_2$ star), we observe a change from  $\lambda_{P_\mathrm{F2}}=1.5875\times10^{-5}+7.8955\mathrm{i}$ to $\lambda_{P_\mathrm{F2}}=-2.8099\times10^{-5}+7.8980\mathrm{i}$. Thereby, compared to fixed SN points, in this case, $\max\text{Re}[\lambda]$  changes sign around critical points: the dynamics turns from oscillatory decay~($\max\text{Re}[\lambda]<0$) to a growth oscillation~($\max\text{Re}[\lambda]>0$) and vice versa, whose population growth is bounded by the nonlinearity of the system indicating that these points, indeed, are related to the Hopf bifurcations. Despite the applicability of the eigenvalue approach to characterize the Hopf bifurcation for increasing pump power, this approach cannot be applied for decreasing pump power.

\begin{figure}
    \centering
    \includegraphics[width=1\linewidth]{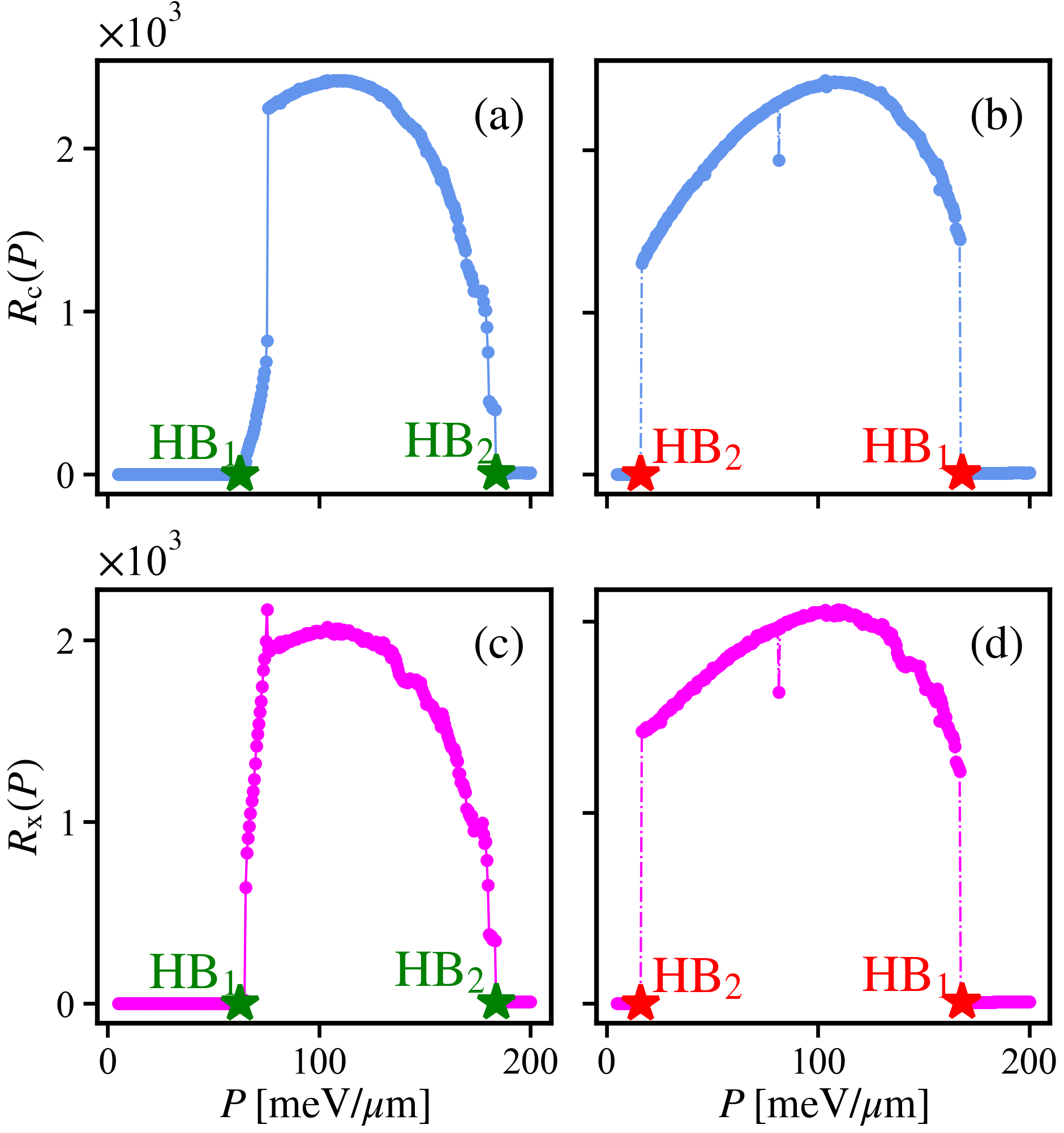}
    \caption{Root-mean-square phase-space radius $R_{\imath}(P)$ of the cavity and exciton populations as a function of pump power $P$ for $\omega=\omega_2$. Panels (a) and (c) show the increasing pump power sweep for the cavity and exciton components, respectively, whereas panels (b) and (d) show the decreasing power sweep. The green stars mark the forward critical powers $P_{\mathrm{HB}_1}\approx62$ and $P_{\mathrm{HB}_2}\approx182$, where the fixed point loses and regains stability through Hopf bifurcations. The red stars indicate the corresponding backward dynamical transition powers, $P\approx167$ and $P\approx16$, delimiting the persistence of the limit-cycle branch during the downward sweep. A vanishing radius identifies a fixed point, whereas a finite radius signals stable limit-cycle oscillations. The mismatch between the forward and downward transition powers demonstrates hysteretic coexistence of fixed-point and limit-cycle attractors.}
    \label{Fig:Two_Component_Power_Population}
\end{figure}

To get a deeper insight into the phenomena and to describe the power evolution of the exciton and polariton during the power sweeps, we also investigate the cavity and exciton populations in the $(|\psi_\imath(t;P)|^2,\partial_t |\psi_\imath(t;P)|^2)$ phase space. Introducing fluctuations about the time-averaged state as $\delta\psi_\imath^2(t;P)^=|\psi_\imath(t;P)|^2-\braket{|\psi_\imath(t;P)|}_t^2$ and $\delta \partial_t|\psi_\imath(t;P)|^2=\partial_t |\psi_\imath(t;P)|^2-\partial_t\braket{|\psi_\imath(t;P)|}_t^2]$, we define the phase-space radius using as the root mean square distance from the time-averaged state
\begin{equation}
    R_\imath(P)=\sqrt{\braket{\delta\psi_\imath^2(t;P)+\delta \partial_t|\psi_\imath(t;P)|^2}_t}.
    \label{Eq:Radius_Two_Component}
\end{equation}
For the increasing sweep, we see a negligible $\text{R}_\imath$ up to the Hopf bifurcation at $P\approx62$, where the distant function starts to grow smoothly, possesses a jump at $P\approx67$, reaches to its maximum at $P\approx105$ and gradually decreases up to $P\approx182$, at which the system relaxes to fixed point through $\text{HB}_2$ Hopf bifurcation; see Fig.~\ref{Fig:Two_Component_Power_Population}(a,c). The non-zero radius of the distant function $\text{R}_\imath$ within $P\in[67,182]$ establishes the excitation of the cavity and exciton population to a stable limit-cycle state bounded by the nonlinearity. The downward evolution shows a transition from a fixed point to a limit cycle at $P\approx 167$~(see red $\text{HB}_1$ stars in Fig.~\ref{Fig:Two_Component_Power_Population}(b,d)), the limit cycle approaches the maximum at $P\approx105$ and relaxes back to a stable fixed point at $P \approx 16$~(red $\text{HB}_2$ stars in Fig.~\ref{Fig:Two_Component_Power_Population}). Interestingly, we discover that the trajectories of the cavity photons and excitons are different for increasing and decreasing power sweeps; in the $P\in[16,67]$, the system has two different attractors: a fixed point~(limit cycle) for increasing~(decreasing) power, which establishes the extension of the concept of hysteresis loop for polaritons excited to a higher limit-cycle states.     

\subsection{Four component model}

\begin{figure}
    \centering
    \includegraphics[width=1\linewidth]{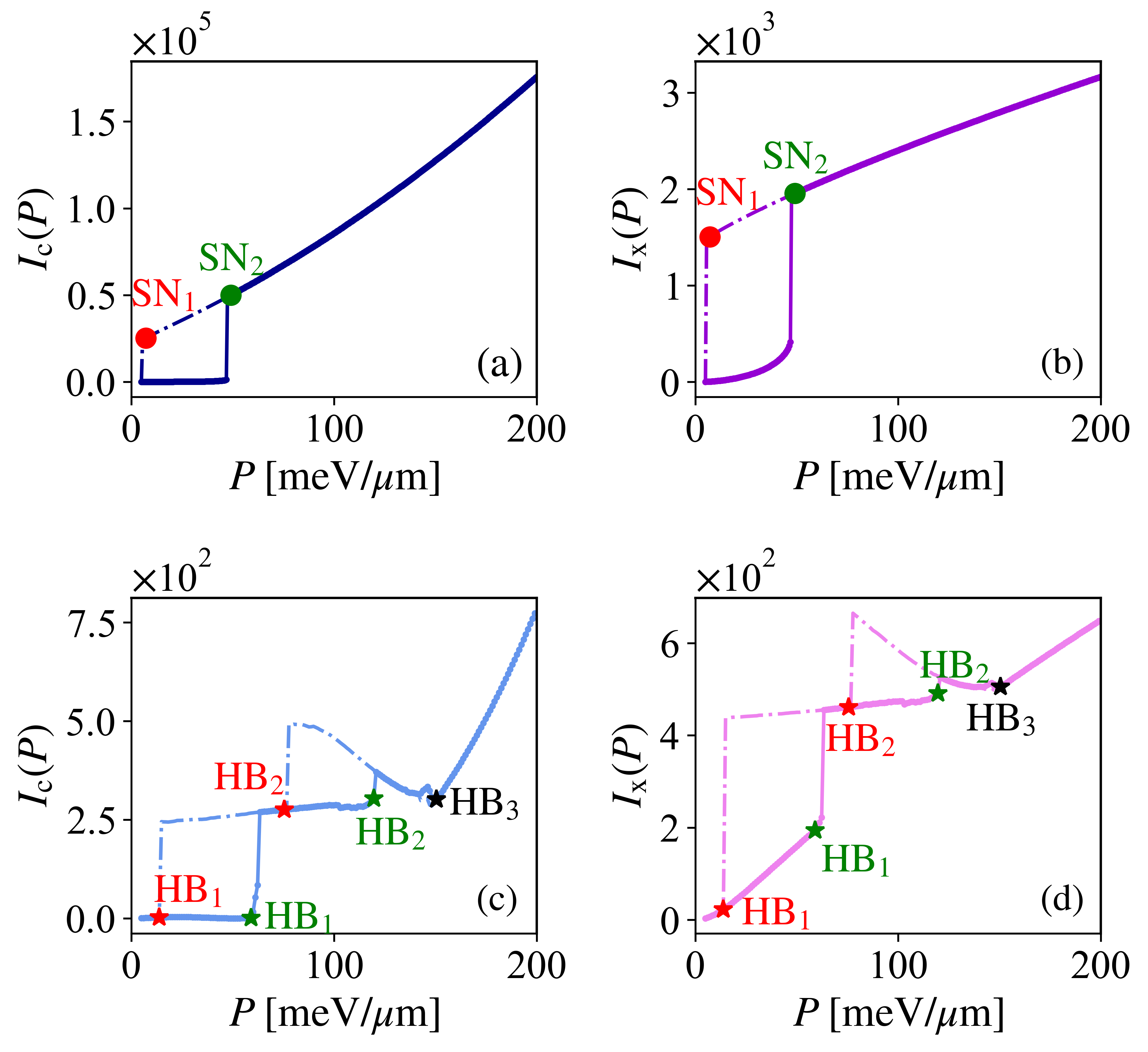}
    \caption{Time-averaged cavity and exciton populations as functions of pump power $P$ for the four-component model. Panels (a) and (b) show $I_{\mathrm{c}}(P)$ and $I_{\mathrm{x}}(P)$, respectively, for $\omega_1=-3$, exhibiting a bistable hysteresis loop bounded by the saddle-node points $\mathrm{SN}_1$ and $\mathrm{SN}_2$. Panels (c) and (d) show the corresponding cavity and exciton populations for $\omega_2=2$. Green and red stars mark the dynamical transition points obtained from the increasing and decreasing power sweeps, respectively, while the black star denotes the high-power transition common to both sweeps. Solid and dash-dotted curves indicate increasing and decreasing power sweeps, respectively.}
    
    \label{Fig_Population_Power}
\end{figure}

We now investigate the four-component dynamics corresponding to $\alpha_2\neq0$ and $\Omega\neq0$. We start our analysis by calculating the exciton and cavity photon populations for both increasing and decreaing power sweeps. Similar to the two-component model, we plot in Fig.~\ref{Fig_Population_Power} $I_\imath(P)$ for the probe frequencies close to the lower polariton and in the mid-gap of the PL spectrum. For $\omega=\omega_1$, excitons and cavity photons are close to the condensation; the interaction does not change the attractor, and for forward and downward pump continuation, we observe bistability and hysteresis loop formation. Similar to the two-component case, we find the critical powers as $P_\mathrm{cr,1}=47.5, P_\mathrm{cr,2}=5$, which is very close to Eq.~\eqref{Eq:Critical_Power_Bistability}. The physical origin is that the four-component PL spectrum for our system parameters and at $k_\mathrm{x}=0$ is very similar to the two-component spectrum. Then we consider the attractor corresponding to these critical pumps as the SN.

For forward pump tuning, at $P \approx 58$, the fixed point loses stability due to dominant inter-exciton-exciton interaction; here, cavity and excitation populations exhibit the first Hopf bifurcation~($\text{HB}_1$, green stars). Similar to a two-component model, the system evolves into a limit cycle. Dissimilarly, as the power increases further, a second Hopf bifurcation occurs at $P \approx 120$~($\text{HB}_2$, green stars), associated with enhanced intra-exciton-exciton nonlinear phase shifts, driving the system into a distinct limit cycle. At higher pump powers, $P \approx 150$, the cavity and pump terms overcome nonlinear phase shifts, which destabilize the limit cycle. 

We now confirm these observations by analyzing the eigenvalues of $\bm J$ for the increasing power sweep; for $\omega=\omega_2$ the two Hopf bifurcations occur at $P_\mathrm{F1}\approx 58$ and $P_\mathrm{F2}\approx152$. To characterize the dynamical features of these jumps, we then calculate the maximum eigenvalues of the Jacobian matrix: for $P_\mathrm{F1}$ maximum eigenvalue change from $\lambda_{P_\mathrm{F1}}=-6.988\times10^{-5}+4.230\mathrm{i}$ to $\lambda_{P_\mathrm{F1}}=2.3209\times10^{-4}+4.230\mathrm{i}$, and for $P_\mathrm{F2}$ we observe a change from  $\lambda_{P_\mathrm{F1}}=1.5527\times10^{-4}+7.2475\mathrm{i}$ to $\lambda_{P_\mathrm{F1}}=-3.7767\times10^{-5}+7.2532\mathrm{i}$. Hence, for $P_\mathrm{F1}$ the attractor changes from a fixed point to a limit cycle by excitation to the higher polarization and excitonic states, while for $P_\mathrm{F2}$, pump power inclusion becomes dominant, the limit-cycle attractor loses its stability and relaxes back to the fixed point. This establishes that the steady-state analysis of the forward pump continuation can be used as a universal framework for characterizing HBs and SNs of the forward pump continuation.

For the decreasing power sweep, the dynamics broadly reverse those of the increasing power sweep; however, nonlinear phase shifts induce a distinct history dependence. In particular, at $P \approx 150$, the fixed point loses stability and, due to the increasing role of the cavity field, undergoes a continuous transition to a limit cycle via spiral trajectories. As the pump power decreases, a stable limit cycle persists down to $P \approx 77$, where inter-exciton-exciton interactions induce a transition to a distinct limit-cycle regime ($\text{HB}_1$, red stars in Fig.~\ref{Fig_Population_Power}). This state remains stable down to $P \approx 14$, where intra-exciton-exciton interactions dominate the nonlinear phase shifts, leading to a second transition~($\text{HB}_2$, red stars) from a stable limit cycle to an attracting spiral trajectory, which subsequently relaxes to a stable fixed point. Both cavity and excitonic wavefunctions remain stable for very-low or very high power, as constant pump power is the only dominant mechanism that governs system evolution.   

\begin{figure}
    \centering
    \includegraphics[width=1\linewidth]{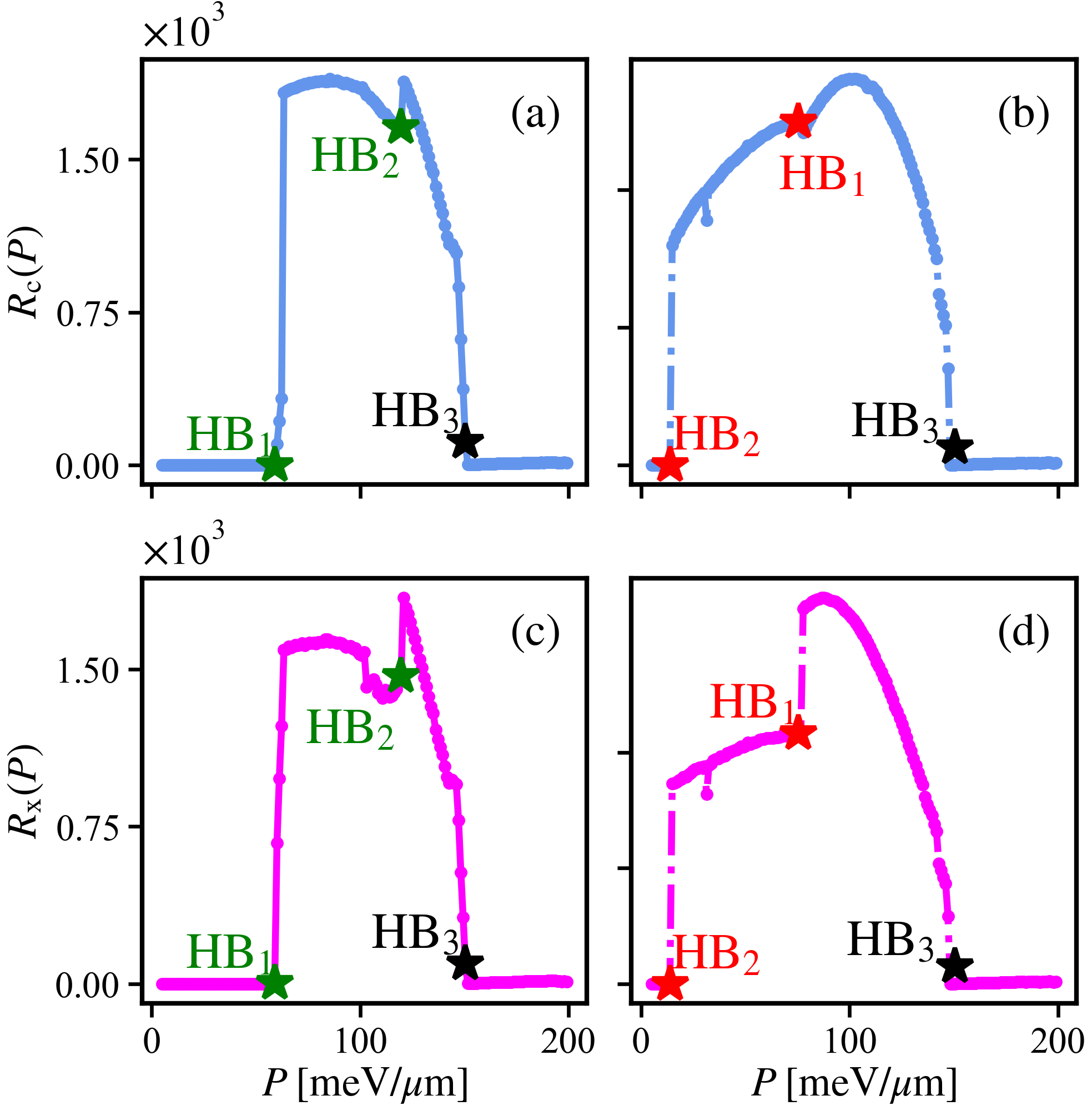}
    \caption{Phase-space trajectory radii of cavity and exciton populations as functions of pump power $P$ for the four-component model. (a) and (c) show $R_{\mathrm{c}}(P)$ and $R_{\mathrm{x}}(P)$ for increasing pump power, respectively, whereas (b) and (d) show for decreasing the pump powr. Green and red stars mark the forward and backward dynamical transition points, respectively, while black stars indicate the high-power transition common to both sweeps. A negligible radius identifies a fixed-point attractor, whereas a finite radius signals limit-cycle dynamics. Different transition powers and intermediate limit-cycle branches for the two sweep directions reveal multiple hysteresis between distinct dynamical attractors.}
    \label{Fig_Oscillation_Power}
\end{figure}

To deepen insight into the attractors of the downward pump continuation, we use the phase space analysis $(I_{\imath}(P),\dot{I}_{\imath}(P))$, where $I_{\imath}(P)=I_{\imath+}(P)+I_{\imath-}(P)$: for each pump $P$, we use Eq.~\eqref{Eq:Power_Population} to calculate the populations and their time derivatives for both increasing and decreasing power sweeps, and we exploit Eq.~\eqref{Eq:Radius_Two_Component} to study the trajectory radius of the cavity photons and exciton evolution in Fig.~\ref{Fig_Oscillation_Power}. Similar to the two-component case, we check the validity of this analysis by considering the increasing pump: we observe a small trajectory radius for both photons and exciton populations up to $P\approx47$ for which $R_\imath(P)$ starts to increase. The trajectory radius then possesses the same cycle as the two-component up to $P \approx 120$ where the second HB happens, and it relaxes to zero value for $P\approx150$. This analysis matches well with the eigenvalue description of the incresing power sweep.

For the decreasing pump power, we find a bifurcation~(transition of the attractor from fixed-point to a limit cycle) at $P\approx150$ for both cavity photon and exciton population, then $R_\imath(P)$ evolves similarly to the downward pump of the two-component case up to $P \approx 120$, where the attractor changes from limit-cycle to another limit-cycle, and the radius decreases up to $P=15$ where the limit-cycle attractor loses its stability, radius becomes negligible and we observe a fixed-point formation. Due to the different classical memory and the formation of different limit cycles, the Hopf bifurcations become different for increasing and decreasing power sweeps; hence, we observe a \textit{multiple} hysteresis loop of the limit cycles for each limit-cycle attractor.     

\section{Synchronization}

As discussed in Sec.~\ref{Sec:Saddle_Hopf}, power tuning of the exciton and cavity photons yields various attractors: close to lower polariton band edge~($\omega=\omega_1$) we achieve saddle node bifurcations, while inside the band gap ($\omega = \omega_2$) Hopf bifurcations appear, which can be explained by eigenvalue and trajectory radius analysis. Despite agreement with numerical solutions of Eqs.~\eqref{Eq:Sudospin_Dynamics_cavity} and \eqref{Eq:Sudospin_Dynamics_exciton}, there are jump-like points in the power spectrum, such as jumps near $P\approx75$ in Fig.~\ref{Two_Component_Power_Population},
which the change of attractors cannot explain. For these points, the attractor is a limit cycle and does not change due to the jump in the power spectrum. Qualitatively, these jumps should be related to nonlinear excitonic phase shifts that act as a saturation of a limit-cycle orbit; quantitatively, these jumps change the limit cycle of the cavity and exciton populations to other limit-cycle states. Through underpinning the theory based on the Floquet multiplier and numerical confirmation by the fast-Fourier transform analysis, we prove that synchronization is the origin of these anomalous jumps.  

To underpin a perturbation theory, we set $k_x=0$, neglect spatial variation and assume  $\psi_{\imath\jmath}(t)=\text{Re}[\psi_{\imath\jmath}(t)]+\mathrm{i}\text{Im}[\psi_{\imath\jmath}(t)]$ to separate the real and imaginary parts of Eqs.~\eqref{Eq:Sudospin_Dynamics_cavity} and \eqref{Eq:Sudospin_Dynamics_exciton}. Introducing $\text{Re}[\psi_{\imath\jmath}(t)]:=\psi_{\imath\jmath}^{\mathrm{r}}$ and $\text{Im}[\psi_{\imath\jmath}(t)]:=\psi_{\imath\jmath}^{\mathrm{i}}$ as short-hand notations, and defining the state vector as
\begin{equation}
    \bm{u}(t)=\left(\psi_\mathrm{c+}^{\mathrm{r}},\psi_\mathrm{c+}^{\mathrm{i}},\psi_\mathrm{c-}^{\mathrm{r}},\psi_\mathrm{c-}^{\mathrm{i}},\psi_\mathrm{x+}^{\mathrm{r}},\psi_\mathrm{x+}^{\mathrm{i}},\psi_\mathrm{x-}^{\mathrm{r}},\psi_\mathrm{x-}^{\mathrm{i}}\right),
    \label{Eq:State_Vector_Synch}
\end{equation}
equations \eqref{Eq:Sudospin_Dynamics_cavity} and \eqref{Eq:Sudospin_Dynamics_exciton} can be written as a vectorial form
\begin{equation}
    \dot{\bm{u}}(t)=\mathcal{F}(\bm{u}).
    \label{Eq:General_Perturbation}
\end{equation}
Considering a limit-cycle solution \(u_\mathrm{LC}(t)\) with initial state $\bm{u}_0=\bm{u}_\mathrm{LC}(0)$ satisfying $\bm{u}_\mathrm{LC}(t+T)=\bm{u}_\mathrm{LC}(t)$; for $T$ the period of the orbit, the nearby trajectory in the phase space can be written as
\begin{equation}
    \bm{u}(t)=\bm{u}_\mathrm{LC}(t)+\delta\bm{u}(t).
    \label{Eq:Perturbation_Limit_Cycle}
\end{equation}
Substituting Eq.~\eqref{Eq:Perturbation_Limit_Cycle} into Eq.~\eqref{Eq:General_Perturbation} and using $\dot{\bm{u}}_\mathrm{LC}(t)=\mathcal{F}(\bm{u}_\mathrm{LC})$, we obtain
\begin{equation}
    \dot{\delta\bm{u}}(t)=\bm{\mathcal{J}}(t)\delta\bm{u}(t),
    \label{Eq:Perturb_Evolution}
\end{equation}
 with matrix entries $\mathcal{J}_{ij}=\partial\mathcal{\bm{F}}_{i}(\bm{u})/\partial u_{j}$ evaluated at $\bm{u}=\bm{u}_\mathrm{LC}(t)$. We also express $\delta\bm{u}(t)$ in terms of the ordinary $\bm{u}_\mathrm{LC}(0)$ and perturbed $\bm{u}_\mathrm{LC}(0)+\delta\bm{u}(0)$ trajectories as 
 \begin{equation}
   \delta\bm{u}(t)=\bm{u}(t,\bm{u}_\mathrm{LC}(0)+\delta\bm{u}(0))-\bm{u}(t,\bm{u}_\mathrm{LC}(0)),  
 \end{equation}
 which upon linearizing the system, and defining $\mathcal{M}_{ij}(t):=\partial_{u_{j}(0)}u_{i}(t,\bm{u}(0))$~[evaluated at \(\bm{u}(0)=\bm{u}_\mathrm{LC}(0)\)] can be written as
 \begin{equation}
     \delta\bm{u}(t)=\bm{\mathcal{M}}(t)\delta\bm{u}(0),
     \label{Eq:Perturbation_M_Matrix}
 \end{equation}
with $\mathcal{M}_{ij}(0)=\delta_{ij}$ as the initial condition. Finally, substituting Eq.~\eqref{Eq:Perturbation_M_Matrix} into Eq.~\eqref{Eq:Perturb_Evolution} we obtain
\begin{equation}
    \dot{\bm{\mathcal{M}}}(t)=\bm{\mathcal{J}}(t)\bm{\mathcal{M}}(t).
    \label{Eq:Dynamic_M_Matrix}
\end{equation}

Computationally, we first use \eqref{Eq:Dynamic_M_Matrix} and calculate the $\bm{\mathcal{M}}(t)$ matrix and plug it into \eqref{Eq:Perturbation_M_Matrix} to find the perturbed wavefunction. To solve \eqref{Eq:Dynamic_M_Matrix}, we make $\bm{\mathcal{J}}(t+T)=\bm{\mathcal{J}}(t)$ as a sole assumption of our methodology. After a period $T$, one can write monodromy matrix $\bm{\mathcal{M}}(T)$ as a eigenfunction $\bm{\mathrm{v}}_j$ and eigenenergies $\mu_j$ as
\begin{equation}
    \bm{\mathcal{M}}(T)\bm{\mathrm{v}}_j=\mu_j\bm{\mathrm{v}}_j.
    \label{Eq:Eigen_M_Matrix}
\end{equation}
Therefore $\bm{\mathrm{v}}_j$ and $\mu_j$ characterize the direction of the perturbation and its Floquet multipliers, respectively. We then expand the arbitrary perturbation $\delta\bm{u}(0)$ in the $\bm{\mathrm{v}}_j$ basis as
\begin{equation}
    \delta\bm{u}(0)=\sum_ja_j\bm{\mathrm{v}}_j.
\end{equation}
Then we obtain the final state of the perturbation as $\delta\bm{u}(mT)=\sum\mu_j^ma_j\bm{\mathrm{v}}_j$. The trivial Floquet multiplier $\mu_0=1$ reflects the neutral time-translation mode of the collective limit cycle. We use this, along with frequency locking $\omega_{\mathrm c}=\omega_{\mathrm x}$ and phase locking 
$\delta\theta\sim\arg[\psi_{\mathrm c}(t)\psi_{\mathrm x}^*(t)]$, to demonstrate synchronized cavity--exciton oscillations.

\begin{figure}
    \centering
    \includegraphics[width=1\linewidth]{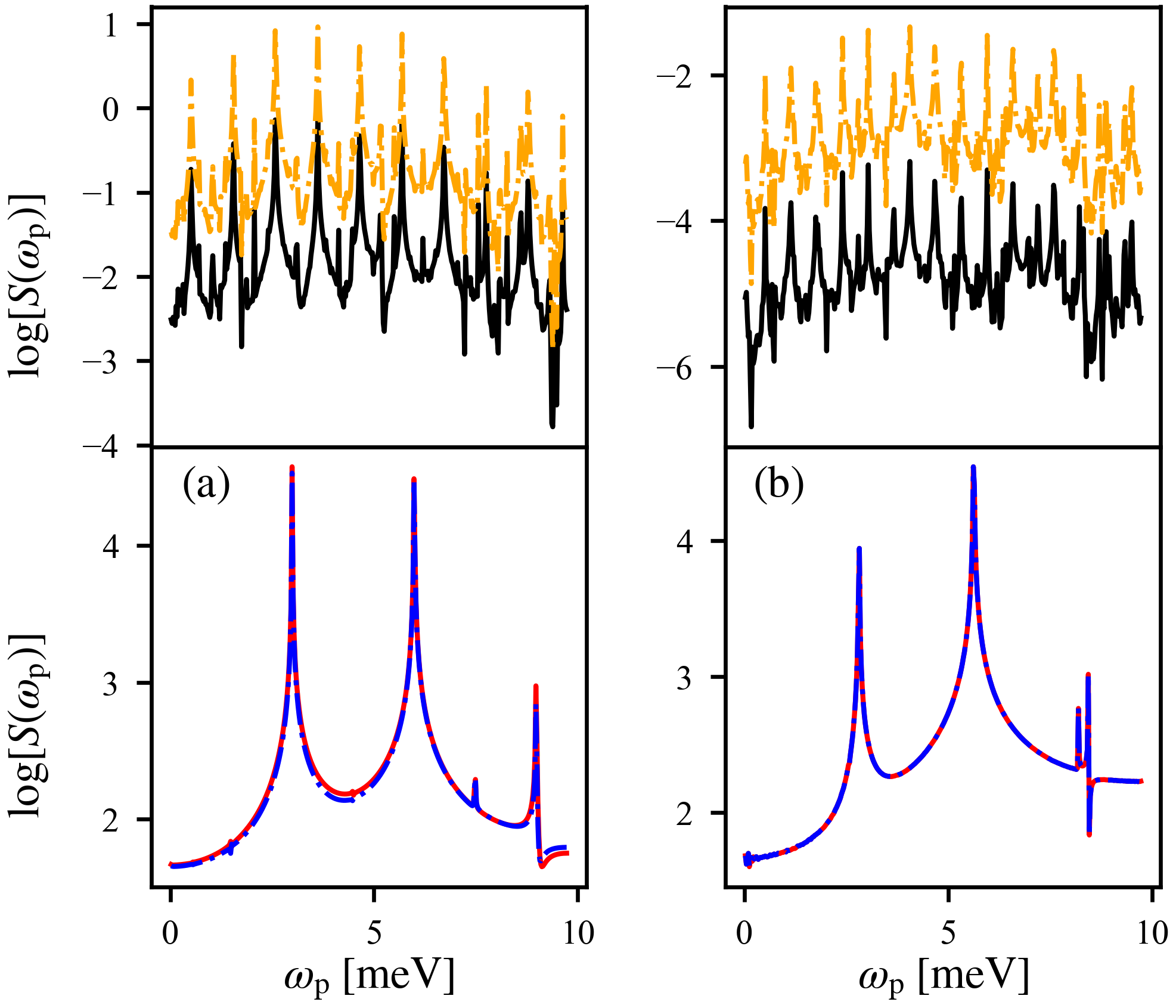}
    \caption{Logarithmic power spectra of the cavity and exciton populations across the population jumps for the (a) increasing- and (b) decreasing power sweeps. The upper panels show the spectra before the corresponding jump, where multiple sidebands and appreciable relative-phase fluctuations indicate weak frequency-comb locking. The lower panels show the spectra after the jump, where the cavity and exciton peaks strongly overlap and only a few dominant sidebands remain, demonstrating enhanced frequency and phase locking. In panel~(a), the spectra are evaluated at $P_{\mathrm{be}}=75.4$ and $P_{\mathrm{af}}=75.9$. The dominant peaks occur near $|\omega_{\mathrm{p}}|=3$, $6$, and $9$, consistent with nonlinear wave mixing and the interaction-induced frequency shift. In panel~(b), the pre-jump downward-pump state is evaluated at $P=167$; after the jump, the coincident spectra are accompanied by vanishing relative-phase fluctuations, indicating complete synchronization of the cavity and exciton populations.}
    \label{Fig_Spectrum}
\end{figure}

\subsection{Two component limit}

We now test the feasibility of our approach for investigating the jump that appears around $P=75$ in Fig.~\ref{Two_Component_Power_Population}(c,d). For $P_\mathrm{be}=75.4$ and $P_\mathrm{af}=75.9$, we find $\mu_0=0.995$ and $\mu_0=0.997$, respectively, indicating that both cavity and exciton populations oscillate collectively. Also, the orbit frequencies of cavity and exciton are locked at $\omega_\mathrm{or}^\mathrm{c}=\omega_\mathrm{or}^\mathrm{x}=4$, suggesting that these states might be synchronized. However, by looking at the logarithmic power spectrum before~(see the upper panel of Fig.~\ref{Fig_Spectrum}(a)) the jump, we observe that the exciton and cavity populations contain multiple sidebands. After the jump~(lower panel Fig.~\ref{Fig_Spectrum}(a)), there are only a couple of sidebands excited with a strong overlap. Thereby, despite the synchronized oscillation of the exciton and cavity photons before and after the jump, jumps cause strongly locked frequency combs.    

We also denote an abnormal phase change through this strong synchronization: before the jump and at $P=74$, the relative phase shows fluctuation $\delta\theta = 0.2\pi$, indicating that the phases are not locked, while after the jump $\delta\theta = 0$ and both populations become synchronized. The logarithmic spectrum also shows the multiple peaks excited due to the wave-mixing of the cavity and exciton populations due to the exciton-exciton interaction. Here, we have a strong pump at $|\omega_\mathrm{c}|=3$, and the nonlinear red-shift detuning causes an excitation of the sidebands at $|\omega_\mathrm{p}|=6$ and $|\omega_\mathrm{p}|=9$.

We now test these considerations for the decreasing power sweep. We consider $P=167$ for this specific case, as is clearly shown in the upper panel of Fig.~\ref{Fig_Spectrum}(b), the relative phases exhibit strong fluctuations $\delta\theta = 0.4\pi$, leading to de-synchronization of the exciton and polariton populations. After the jump, however, we recover $\delta\theta = 0$ and observe a perfect synchronization of the system. 

\subsection{Four component model}

We now investigate the synchronization for the four-component case and for both forward and downward pump continuation. For increasing power, we observe discontinuity of power spectrum at $P\approx62$~(see Fig.~\ref{Fig_Spectrum}(a,e) for cavity and exciton spectrum, respectively) and $P\approx120$~(indicated at Fig.~\ref{Fig_Spectrum}(b,f) for cavity and exciton spectrum, respectively), while for decreaing power a discontinuity occurs for $P\approx75$~(see Fig.~\ref{Fig_Spectrum}(c,g)) and $P\approx15$~(see Fig.~\ref{Fig_Spectrum}(d,h) for power spectrum). In what follows, we discuss the forward and downward cases separately. 

For the forward pump continuation, we employ the Floquet formalism commensurate with the frequency-locking and phase fluctuation to study the synchronization among exciton and cavity photons. For $P\approx63$, we achieve $|\mu_0|=1$, while the phase of the cavity photons and excitons before $P=62$ and after $P=64$ jumps by $\pi$. Furthermore, we obtain phase fluctuations for $P=62$ as $\delta\theta = 0.15\pi$ while for  $P=64$, $\delta\theta = 0$ indicating a perfect synchronization. On the other hand, around $P\approx120$ we achieve $|\mu_0|=1$, and observe the same phase jump $\pi$ before and after the jump. Then, for $P\approx119$ we find $\delta\theta = 0.04\pi$, while for $P\approx120$ we observe a perfect synchronization. These considerations justify that the state of the excitons and cavity photons before and after jumps are synchronized; however, the jumps cause a minimum in phase fluctuations, yielding a perfect synchronization in the system.

To gain a deeper insight into the origin of the jumps, we next discuss the logarithmic power spectrum of the components as well as total cavity photon-exciton evolution in Fig.~\ref{Fig_Spectrum}. Since only the positive cavity component is pumped by external fields, positive components of the cavity and exciton spectra~(shown by orange and red dashed-dotted lines in Fig.~\ref{Fig_Spectrum}) share similar features to the total power spectrum of the cavity and exciton~(represented by black and blue solid lines in Fig.~\ref{Fig_Spectrum}). Before the first forward discontinuity $P=62$~[Figs.~\ref{Fig_Spectrum}(a) and \ref{Fig_Spectrum}(e)], positive and negative components of the cavity photons evolve dissimilar to the corresponding exciton components, yielding a noticeable phase fluctuation~[see upper panels of Figs.~\ref{Fig_Spectrum}(a) and \ref{Fig_Spectrum}(e)]. After jump $P=64$, we observe perfect matching between the spectral components of the cavity photons and excitons~(as shown in the lower panels of Figs.~\ref{Fig_Spectrum}(a) and \ref{Fig_Spectrum}(e)). The same analysis is valid for the jump that occurred at $P=120$; however, for this case, negative cavity and exciton components become dominant~(see strong variations of the dashed magenta and green lines around $\omega_\mathrm{p}=9$~meV that appear in Figs.~\ref{Fig_Spectrum}(b) and \ref{Fig_Spectrum}(f), respectively); before jump, the nonlinear phase shift due to this spectral evolution yield a phase fluctuation in the total spectrum which becomes a stable frequency combs after perfect synchronization.        

Despite the inaccuracy of the Floquet approach for the downward pump, the outlined spectral analysis represents a powerful approach for investigating the origin of the jumps for this driven system. For the jump that occurred at $P=78$, negative components become dominant~(as clearly shown in Figs.~\ref{Fig_Spectrum}(c) and \ref{Fig_Spectrum}(g), respectively); the system possesses a phase lag due to this nonlinear shift, which causes a phase fluctuation in the cavity and exciton component; through the jump, the system become perfectly synchronized and the nonlinear phase shift evolve into the stable frequency comb. The interacting system evolves as perfectly synchronized states up to $P=15$~(see bottom panels of Figs.~\ref{Fig_Spectrum}(d) and \ref{Fig_Spectrum}(f), respectively), for which the system changes the attractor and relaxes to a fixed-point attractor; the frequency combs lose their stability and evolve to the weak broadband noise~(see upper panels of Figs.~\ref{Fig_Spectrum}(d) and \ref{Fig_Spectrum}(f) for $P=10$, respectively).

\begin{figure*}
    \centering
    \includegraphics[width=1\linewidth]{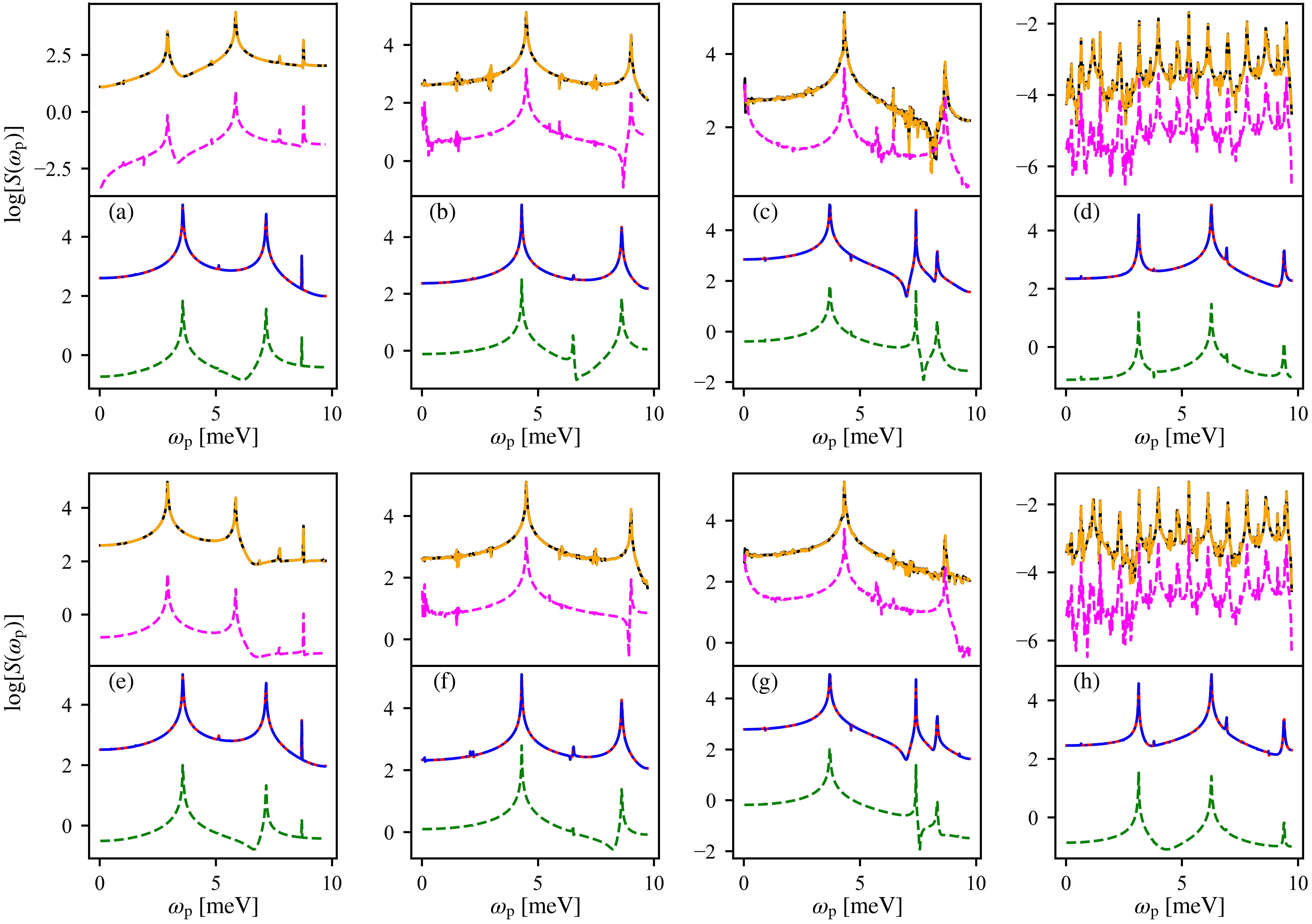}
    \caption{Logarithmic power spectra of the four-component exciton--polariton system across the discontinuities obtained under forward and downward pump continuation. Panels~(a)--(d) show the cavity spectra, while panels~(e)--(h) show the corresponding exciton spectra. In each panel, the upper and lower subpanels represent the states before and after the associated jump, respectively. The total cavity and exciton spectra are shown by the black and blue solid lines, while the positive-polarization components are indicated by the orange and red dash-dotted lines and the negative-polarization components by the magenta and green dashed lines, respectively. Panels~(a) and~(e) correspond to the first forward discontinuity near $P\simeq62$, where the pre-jump spectra display a clear mismatch between the corresponding cavity and exciton components and a finite phase fluctuation, whereas after the jump the spectral peaks overlap and the system becomes perfectly synchronized. Panels~(b) and~(f) correspond to the second forward discontinuity near $P\simeq120$; here, the negative-polarization components become dominant, producing a nonlinear phase shift and finite phase fluctuation before the jump, while the post-jump state forms a stable, frequency-locked comb. Panels~(c) and~(g) show the downward-pump discontinuity near $P\simeq75$, for which the dominant negative components similarly evolve from a phase-lagged state into a perfectly synchronized frequency comb. Panels~(d) and~(h) correspond to the lower downward-pump transition near $P\simeq15$: the post-jump oscillatory state exhibits well-defined and mutually locked spectral peaks, whereas the lower-power state relaxes toward a fixed-point attractor and the comb structure is replaced by weak broadband fluctuations. }
    \label{Fig_Spectrum}
\end{figure*}
 
For forward pumping with positive cavity polarization, we observe a clear difference between $\psi_{\mathrm{c}+}$ (black dot–dashed line in Fig.~\ref{Fig_Spectrum}(a)) and $\psi_{\mathrm{x}+}$ (green dot–dashed line) prior to the $\mathrm{HB}_1$ transition. After this Hopf bifurcation, the system evolves into a stable limit cycle, where the positive components become frequency-locked with a well-defined phase difference (blue and red dot–dashed lines for $\psi_{\mathrm{c}+}$ and $\psi_{\mathrm{x}+}$, respectively). Following $\mathrm{HB}_1$, the negative components also become frequency-locked, leading to full synchronization between all cavity and exciton components, consistent with the emergence of a limit cycle observed in the power spectrum [Fig.~\ref{Fig_Population_Power}(a),(b)]. As the pump power approaches $\mathrm{HB}_2$, the positive components remain synchronized; however, $\psi_{\mathrm{c}-}$ and $\psi_{\mathrm{x}-}$ exhibit clear asynchronism (black and green dot–dashed lines in Fig.~\ref{Fig_Spectrum}(b)). This behavior precedes a weaker transition, consistent with $\alpha_2 \ll \alpha_1$, where the system undergoes a second bifurcation into another limit cycle that restores synchronization among all components (see also Fig.~\ref{Fig_Population_Power}(a),(b)).

\section{Conclusion}

We have shown that resonantly driven exciton-polariton microcavities support a rich hierarchy of dynamical states extending beyond conventional optical bistability. While saddle-node bifurcations govern the familiar bistable response near the lower polariton branch, tuning the pump frequency into the regime where excitonic and photonic contributions are comparable gives rise to multiple Hopf bifurcations and stable limit-cycle oscillations. These oscillatory states generate pronounced dynamical hysteresis, with distinct transition powers observed during forward and backward pump sweeps.

In the spinor system, weak polarization splitting and spin-dependent nonlinear interactions further enrich the dynamics, producing multiple limit-cycle branches and hysteresis loops between different dynamical attractors. Linear stability analysis, long-time time-domain simulations, and phase-space diagnostics consistently reveal transitions between stable fixed points and self-sustained oscillatory states. The locations of these transitions can be controlled through the pump frequency and system parameters.

Most importantly, our analysis identifies synchronization of the cavity-photon and exciton degrees of freedom as the organizing principle underlying the observed dynamics. The onset of limit cycles, discontinuous transitions between oscillatory branches, stabilization of frequency combs, and the resulting dynamical hysteresis all coincide with changes in frequency and phase locking between the constituent fields. These results provide a unified framework for understanding nonlinear dynamical phenomena in driven spinor polariton systems and suggest new opportunities for controlling oscillatory states and frequency-comb generation using exciton-polaritons~\cite{PhysRevLett.114.193901,Egorov2018,PhysRevB.101.085302}.

While our model captures the essential nonlinear spinor dynamics responsible for the observed Hopf bifurcations and dynamical hysteresis, it neglects spatial effects that may become important in realistic devices. Future work could investigate how synchronization-driven transitions manifest in spatially extended polariton systems including lattices, where propagation effects, disorder, and topology provide additional degrees of freedom~\cite{Schneider2017}.

\section*{Acknowledgements}

D.L. acknowledges support from the Ministry of Education, Singapore, under its SUTD Kickstarter Initiative (Grant No. SKI 20210501). V.V.K was supported by the Fundação para a Ciência e Tecnologia under the project UID/00618/2025 (DOI https://doi.org/10.54499/UID/00618/2025).

\bibliography{apssamp}

\end{document}